\documentclass[a4paper,12pt]{article}
\usepackage{amsmath}
\usepackage{amssymb}
\usepackage{psfig} 
\usepackage{latexsym}
\usepackage[dvips]{graphicx}  
\usepackage[all]{xy}          
\usepackage{amsfonts}

\begin{document}
\begin{titlepage}
\begin{flushright}
IFUP--TH 2003/46 \\
\end{flushright}
~

\vskip 4truecm
\begin{center}
\Large\bf
The tetrahedron graph in Liouville theory on the pseudosphere 
\footnote{This work is  supported in part
  by M.U.R.S.T.}
\end{center}

\vskip 1truecm
\begin{center}
{Pietro Menotti} \\ 
{\small\it Dipartimento di Fisica dell'Universit{\`a}, Pisa 56100, 
Italy and}\\
{\small\it INFN, Sezione di Pisa}\\
\end{center}
\vskip .8truecm
\begin{center}
{Erik Tonni} \\  
{\small\it Scuola Normale Superiore, Pisa 56100, Italy and}\\
{\small\it INFN, Sezione di Pisa}\\
\end{center}
\begin{center}
November 2003
\end{center}
\end{titlepage}

\begin{abstract}
  \noindent We compute analytically the tetrahedron graph in Liouville 
  theory on the pseudosphere. The result allows to extend the 
  check of the bootstrap formula of Zamolodchikov and Zamolodchikov to
  third order perturbation theory of the coefficient $G_3$. We obtain
  complete agreement.  
\end{abstract}

\noindent A.B. Zamolodchikov and Al.B. Zamolodchikov \cite{ZZ:Pseudosphere}
proposed an exact non perturbative formula for the one-point function
$\left\langle \, V_{a}(z_1) \,\right\rangle= \left\langle \, e^{2a\,
    \phi(z_1)}\,\right\rangle $ in Liouville theory on the Poincar\'e-
Lobachevskiy 
pseudosphere, being $\phi$ the Liouville field. The exact quantum
conformal dimensions of the vertex operator $ V_{a}(z_1)$ 
are $a\,(Q-a)$. The background charge $Q$ is related
to the coupling constant $b$ by $Q=b^{-1}+b$, which is related to the
central charge of the theory $c$ by $c=1+6\,Q^2$. Thus, the form of the
one-point function is
\begin{equation}
 \label{onepoint}
\left\langle \,V_{a}(z_1)\,\right\rangle =
\frac{U(a)}{(\,1-z_1\bar{z}_1\,)^{2a(Q-a)}}
\end{equation}
where the proposed formula for the structure constant $U(a)$ is
\cite{ZZ:Pseudosphere}
\begin{equation}
 \label{U(a)bootstrap}
U(a)=\left[\,\pi\mu\gamma(b^2) \,\right]^{-\frac{a}{b}}\,
\frac{\Gamma(bQ) \; \Gamma(Q/b)\;Q}{\Gamma(b(Q-2a))\;
\Gamma(b^{-1}(Q-2a))\;(Q-2a)}~.
\end{equation}
A similar conjecture has been put forward for the structure constant
of the three point function on the sphere \cite{ZZ:Sphere} and in this
case it has 
been verified up to order $b^{10}$ in perturbation theory \cite{Thorn:PT},
within the hamiltonian approach \cite{CT, BCT:Exact}.  \\
It is useful to consider \cite{ZZ:Pseudosphere} instead of
$\left\langle \, e^{2a\phi(z_1)}\,\right\rangle$  the expansion in
powers of $a$ of the logarithm of (\ref{onepoint})
\begin{equation}
 \label{log(one-point function)}
\log \left\langle \, e^{2a\phi(z_1)}\,\right\rangle  = 
\sum_{n=1}^\infty \frac{(2a)^n}{n!}\,G_n .
\end{equation}
From this definition, we have
\begin{eqnarray}
G_1 &=& \langle \phi(z_1) \rangle \\
G_2 &=& \langle \phi^2(z_1) \rangle -  \langle \phi(z_1) \rangle^2 \\
G_3 &=&  \langle \phi^3(z_1) \rangle -
              3 \langle \phi^2(z_1) \rangle\langle \phi(z_1) \rangle+
              2 \langle \phi(z_1) \rangle^3.
\end{eqnarray}
In \cite{ZZ:Pseudosphere} also a perturbative check
for $U(a)$ was given, by computing $G_1$ to order $b^3$, $G_2$ to
order $b^2$ and
$G_3$ to order $b$. In this letter, we extend the computation of $G_3$
to order $b^3$.\\ 
Using the representation on the unit disk $\Delta=\{\,|z|<1\,\}$ for
the pseudosphere, the Liouville action is written as
\begin{equation}
 \label{action}
S_{\Delta}[ \,\phi\,] = \int_{\Delta}  \left[\,\frac{1}{\pi}
  \partial_z  \phi \, \partial_{\bar{z}}\phi + \mu
  e^{2b\phi}\,\right]\,d^2 z ~.
\end{equation}
Decomposing the field $\phi$ as a sum of a background field
$\phi_{cl}$ and a quantum fluctuation $\chi$
\begin{equation}
 \label{decomposition}
\phi=\phi_{cl}+\chi
\end{equation}
with
\begin{equation}
 \label{phi_cl}
\phi_{cl} (z)= -\,\frac{1}{2b}\,\log \left[ \,\pi b^2 \mu\,
  (1-z\bar{z})^2 \, \right] 
\end{equation}
the action becomes \cite{ZZ:Pseudosphere}
\begin{equation}
S_{\Delta}[ \,\phi\,] = S_{\Delta}[ \,\phi_{cl}\,]+
\int_{\Delta}  \left[\,\frac{1}{\pi} \partial_z  \chi \,
\partial_{\bar{z}}\chi  
+ \frac{e^{2b\chi}-1-2b\chi}{\pi b^2\,(1-z\bar{z})^2}\,\right]\,d^2 z.
\end{equation}
The field $\phi_{cl}$ describes a surface of constant negative
curvature.\\
Decomposing the lagrangian of the quantum field $\chi$ in the
quadratic part and in the interaction part, one gets the propagator
\cite{ZZ:Pseudosphere}
\begin{equation}
 \label{g(z,z')}
g(z,z') =  \,  -\, \frac{1}{2}\, 
\left(\,\frac{1+\,\eta}{1-\,\eta}\, \log \,\eta \, +2\,\right)
\end{equation}
being $\eta(z,z')$ the $SU(1,1)$ invariant
\begin{equation}
 \label{eta(z,z_n)}
\eta (z,z')= \left|\,\frac{z-z'}{1-z\bar{z}'}\,\right|^2.
\end{equation}
The interaction lagrangian is given by
\begin{equation}
\int_\Delta \frac{d^2 z}{\pi b^2\,(1-z \bar z)^2}\,\sum_{k=3}^{\infty}
\frac{(2 b)^k}{k!}\,\chi^k.
\end{equation}
The graphs contributing to the order $b^3$ in $G_3$ are shown below\\
\phantom{-----------------}
\[
\vspace{1cm}
\begin{array}{ccc}
\vspace{1cm}
\textrm{(a)}\hspace{0.5cm}
\begin{minipage}[c]{3.5cm} 
  \includegraphics[width=2.1cm]{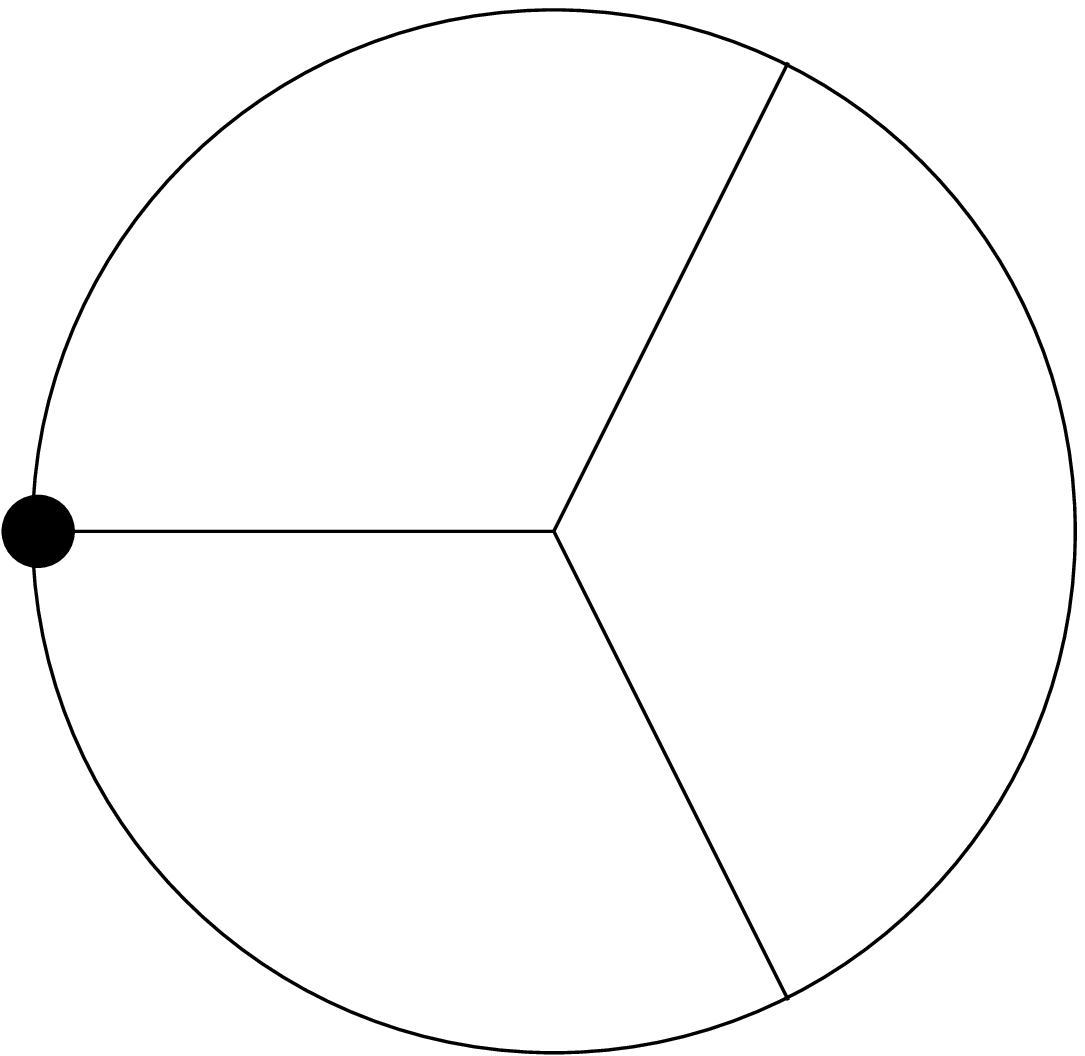}
\end{minipage} 
&
\hspace{1cm}
\textrm{(b)}\hspace{0.5cm}
\begin{minipage}[c]{3.5cm} 
  \includegraphics[width=3.2cm]{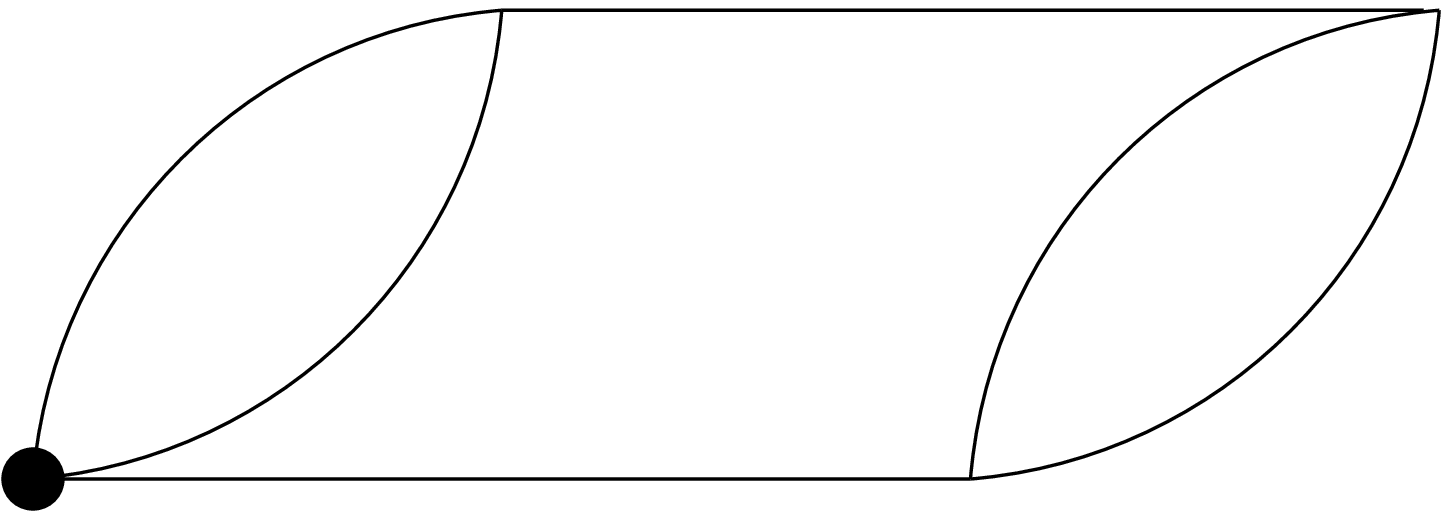}
\end{minipage} 
&
\hspace{1.8cm}
\textrm{(c)}\hspace{0.5cm}
\begin{minipage}[c]{3.5cm} 
  \includegraphics[width=2cm]{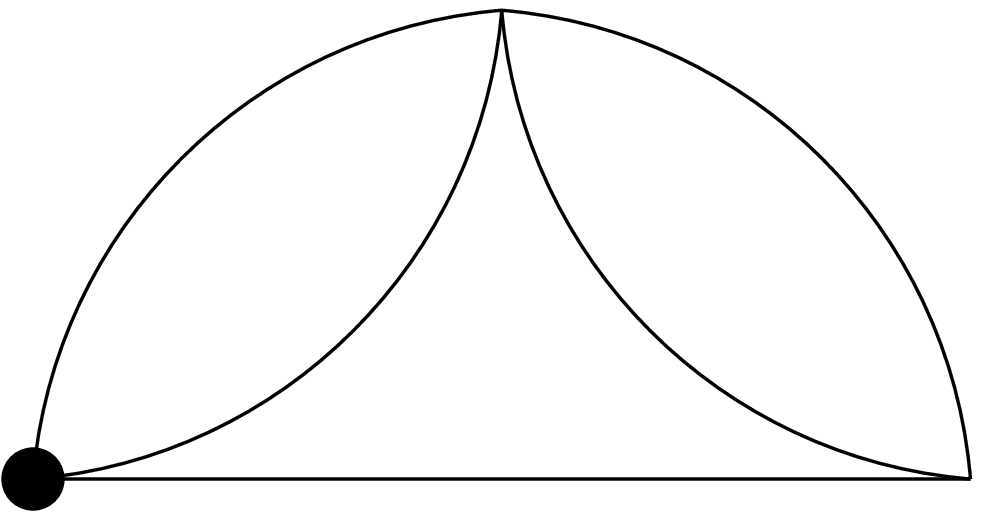}
\end{minipage} 
\\
\vspace{.7cm}
\textrm{(d)}\hspace{0.5cm}
\begin{minipage}[c]{3.5cm} 
  \includegraphics[width=3.3cm]{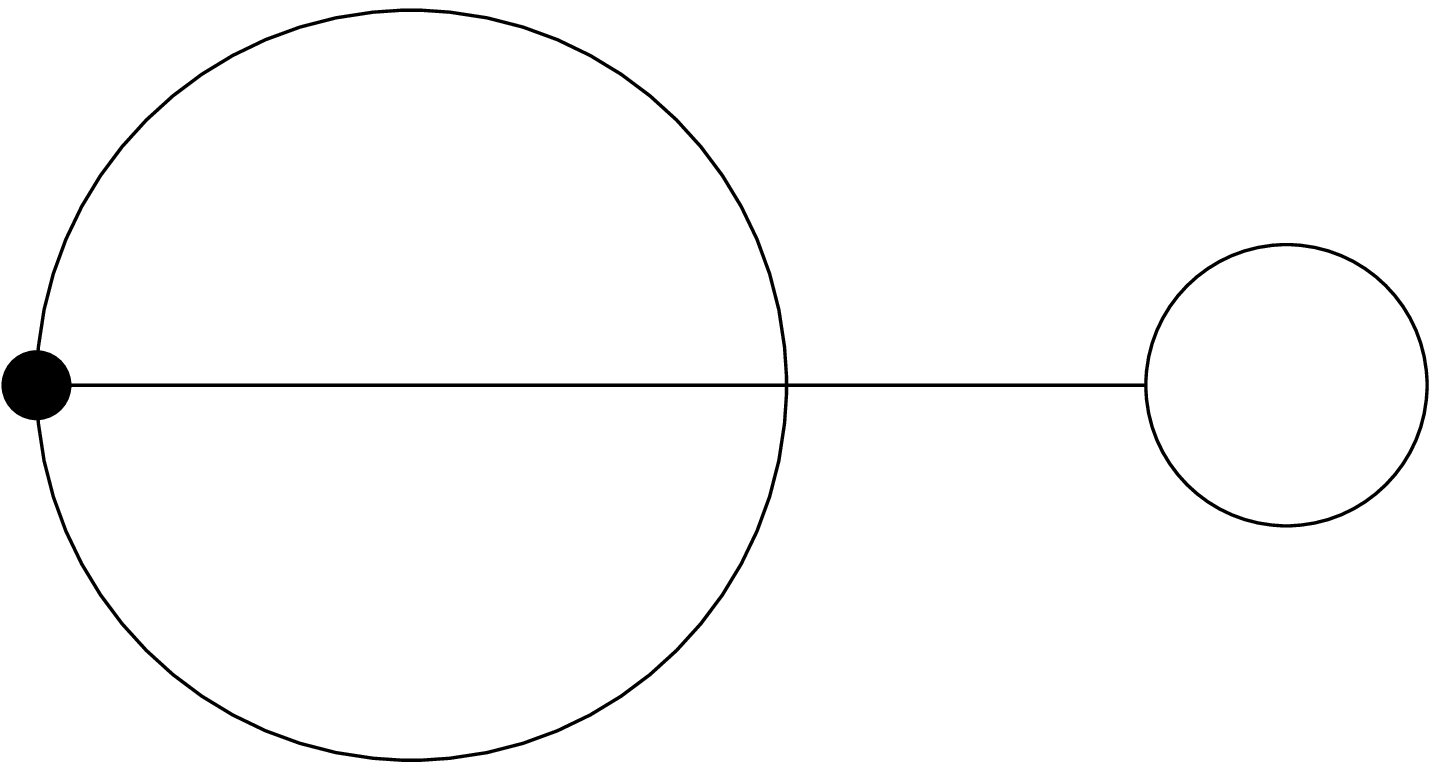}
\end{minipage} 
&
\hspace{1cm}
\textrm{(e)}\hspace{0.5cm}
\begin{minipage}[c]{3.5cm} 
  \includegraphics[width=2.5cm]{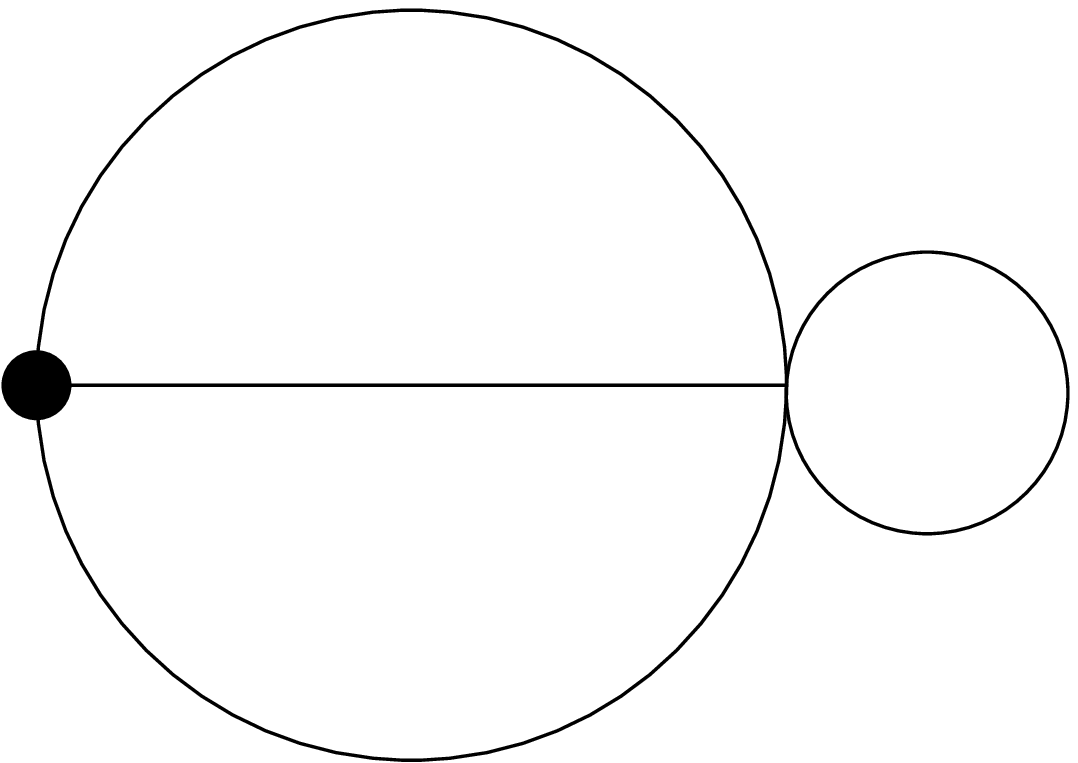}
\end{minipage} 
&  
\end{array}
\]
\[
\vspace{0.7cm}
\begin{array}{ccc}
\textrm{(f)}\hspace{0.5cm}
\begin{minipage}[c]{3.5cm} 
  \includegraphics[width=1.9cm]{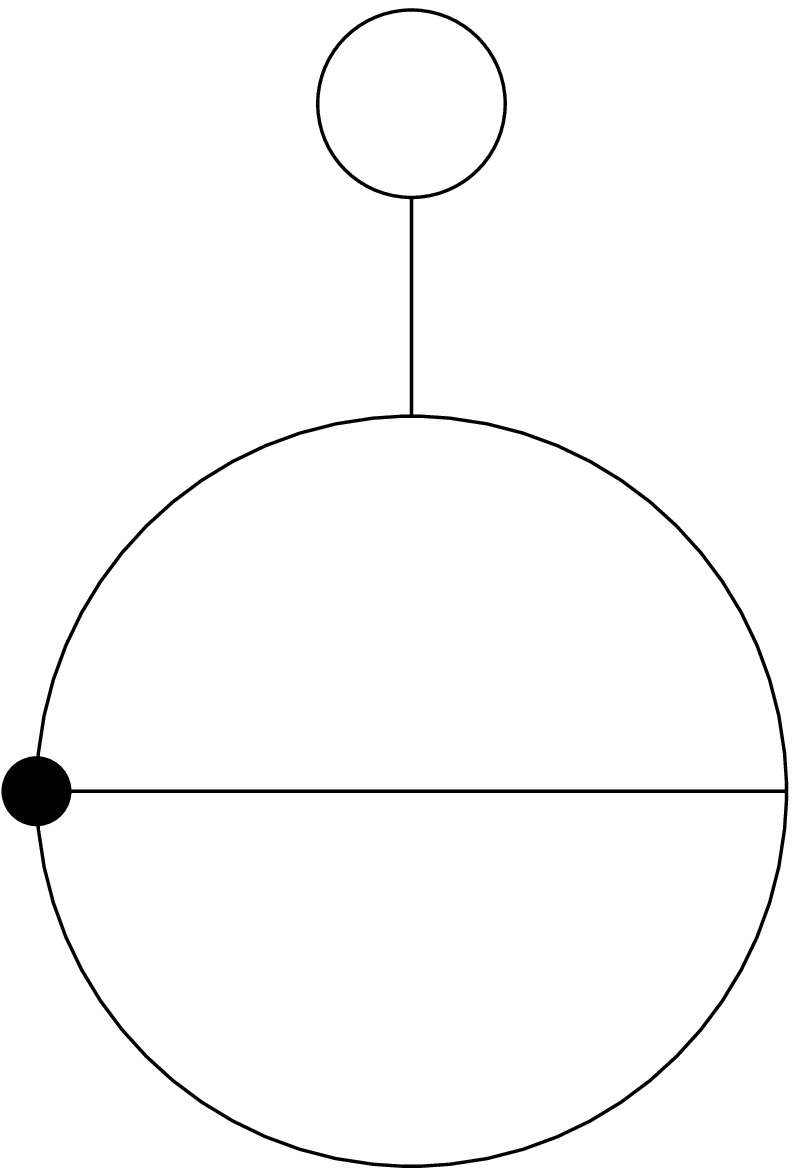}
\end{minipage} 
&
\hspace{1.1cm}
\textrm{(g)}\hspace{0.5cm}
\begin{minipage}[c]{4.5cm} 
  \includegraphics[width=1.9cm]{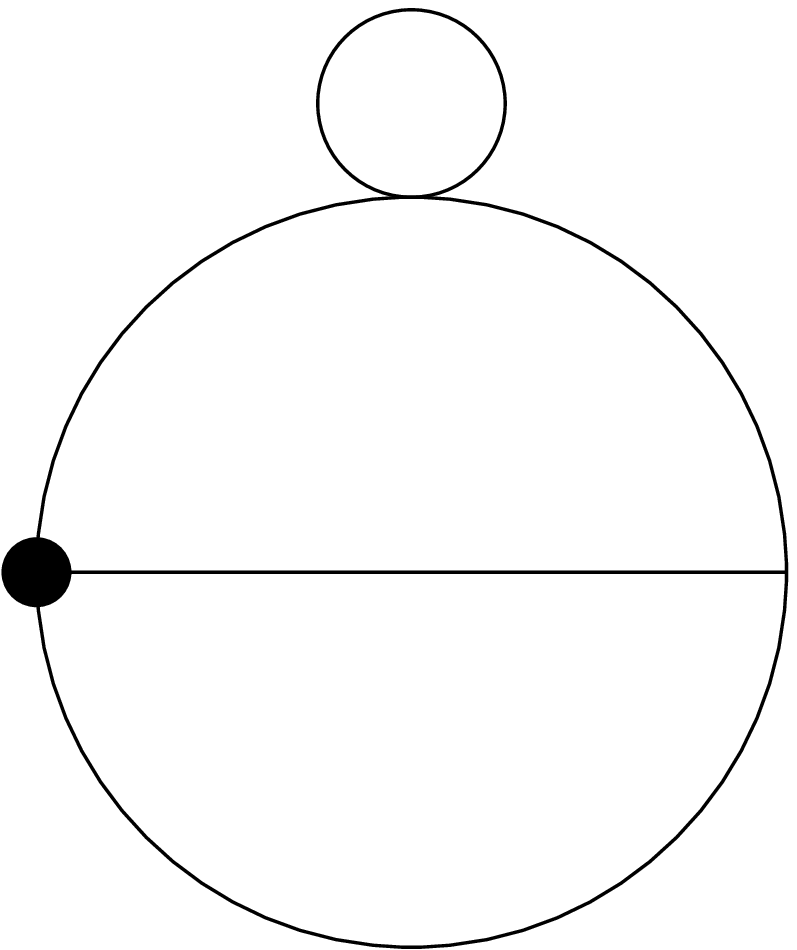}
\end{minipage} 
& 
\end{array}
\]
The most difficult graph is the first one, known as the tetrahedron
(or mercedes) graph and we will present a technique to calculate it. 
Before doing that, we will discuss the simpler graphs (b), \dots , 
(g).\\
Graph (b) is obtained by integrating on the invariant measure
$d^2z/\pi (1-z\bar z)^2$ the square of the subgraph shown below and
already computed in \cite{ZZ:Pseudosphere} 
 
\[
\vspace{.5cm}
\begin{minipage}[c]{3.2cm} 
  \includegraphics[width=3.3cm]{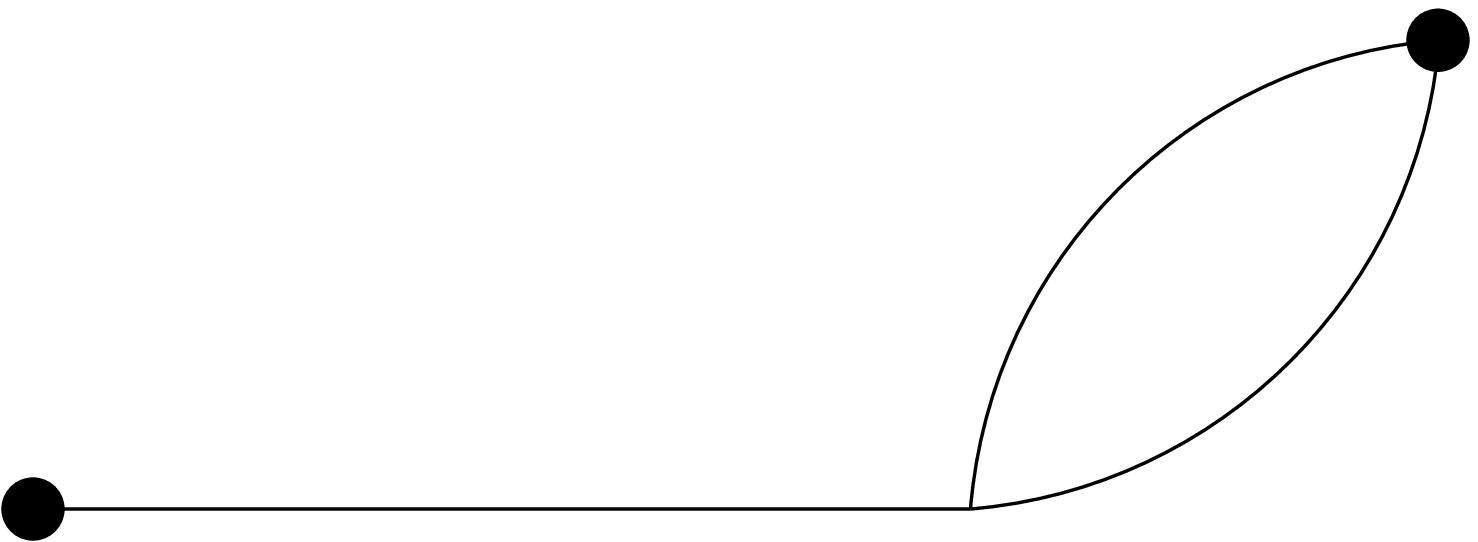}
\end{minipage} 
\hspace{.2cm} = \hspace{.2cm} \frac{1}{\pi}\,g_{1,2}(z,z')
\hspace{.2cm} = 
\hspace{.2cm} \int_{\Delta}
\frac{g(z,z'')\,g^2(z'',z')}{\pi\,(1-z''\bar{z}'')^2} \hspace{.2cm} =
\hspace{.2cm} -\,\frac{1}{8}\left(\,\frac{\eta
    \log^2\eta}{(1-\eta)^2}-1\,\right).
\]
Graph (c) can be also reduced to an integration of the previous
subgraph and 
the square of $g(z,z')$ on the invariant measure. Graphs (d), (e), (f)
and (g) contain the divergent loop. Following the 
regularization proposed in
\cite{ZZ:Pseudosphere}, we will put
\begin{equation}
g(z,z)\equiv\lim_{z'\rightarrow z}\Big(\,g(z,z')+\log|z-z'|\,\Big) = 
\log\,(1-z\bar
z)-1.
\end{equation}
The sum of the graphs (d) and (e) collapses into graph (j) shown below,
where the cross means the introduction of the vertex $b^3 \chi^3$.\\
\phantom{--------------------}
\begin{equation}
\vspace{0.3cm}
\begin{array}{cccccc}
\vspace{.5cm}
\begin{minipage}[c]{3.2cm} 
  \includegraphics[width=3cm]{G3-3C.eps}
\end{minipage} 
& \hspace{0.4cm} +  \hspace{0.5cm}&
\begin{minipage}[c]{3.2cm} 
  \includegraphics[width=2.3cm]{G3-3C1.eps}
\end{minipage}
& \hspace{-0.5cm} = & 
\hspace{0.5cm}
\begin{minipage}[c]{3.2cm} 
  \includegraphics[width=1.8cm]{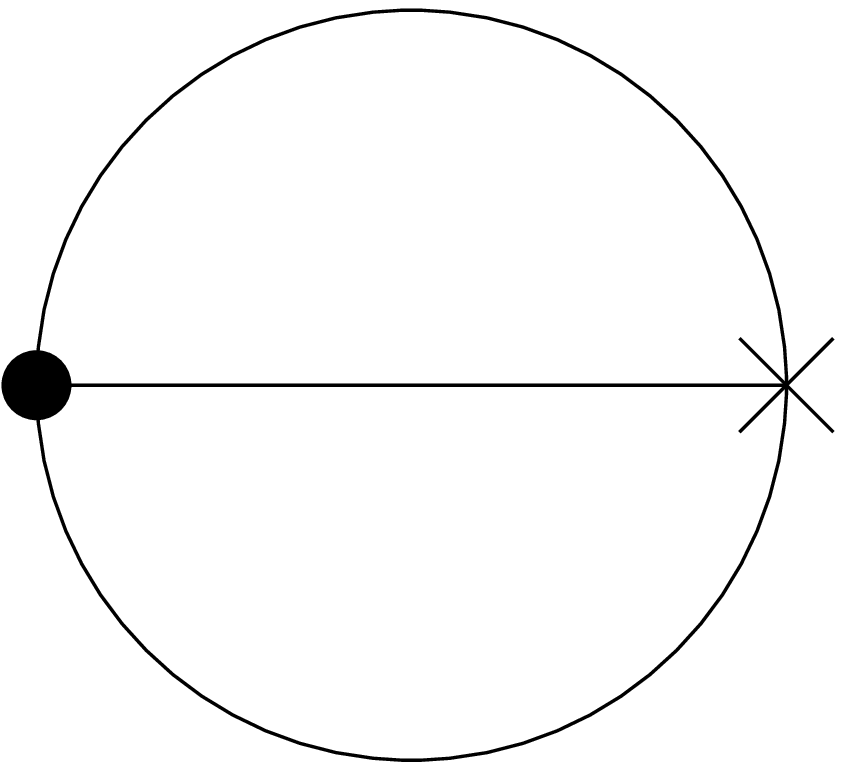}
\end{minipage}
\hspace{-0.7cm}\textrm{(j)}
\end{array}
\end{equation}
Similarly, the sum of (f) and (g) collapses into graph (k), where the
cross means an interaction vertex $b^2\chi^2$.\\
\phantom{------------------------}
\begin{equation}
\begin{array}{cccccc}
\begin{minipage}[c]{3.2cm} 
  \includegraphics[width=1.8cm]{G3-3D.eps}
\end{minipage}
& \hspace{-0.6cm} + \hspace{0.7cm} &
\begin{minipage}[c]{3.2cm} 
  \includegraphics[width=1.8cm]{G3-3D1.eps}
\end{minipage}
& \hspace{-0.7cm} = & 
\hspace{0.7cm}
\begin{minipage}[c]{3.2cm} 
  \includegraphics[width=1.8cm]{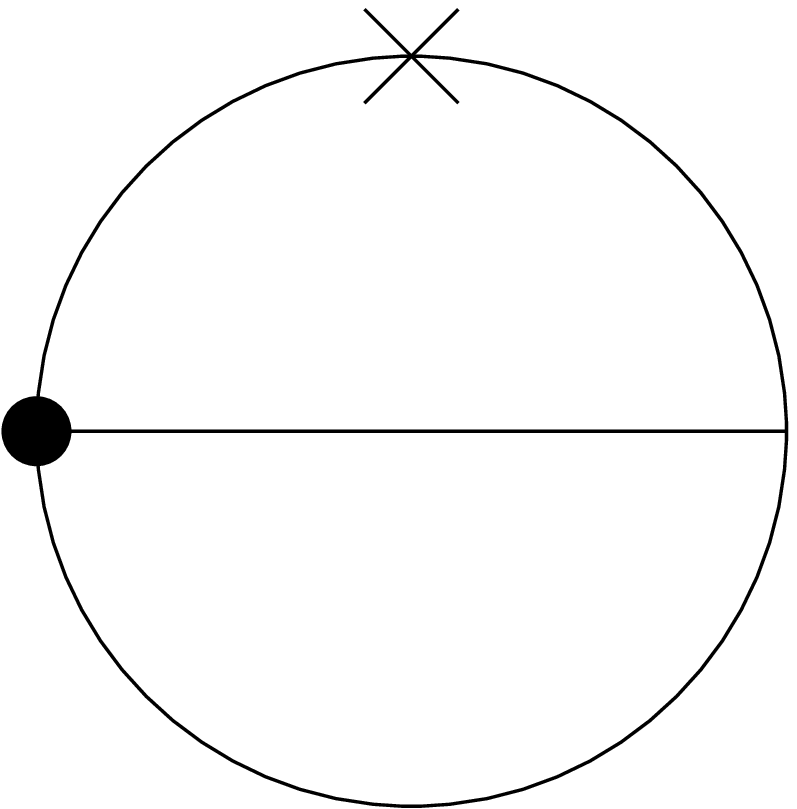}
\end{minipage}
\hspace{-0.7cm}\textrm{(k)}
\end{array}
\end{equation}
Summing up, taking into account the correct multiplicities, we obtain
\begin{eqnarray}
\textrm{(b)}&=&\left(\,\frac{16}{5}+\frac{2}{3}\,\pi^2-\frac{8}{75}\,\pi^4\,
\right)b^3
 \nonumber\\
\textrm{(c)} &=& \left(\, -\frac{18}{5}+\frac{4}{75}\,\pi^4 \,\right)b^3
 \nonumber\\
\textrm{(j)} &=& \left(\, -1 \,\right)b^3
 \nonumber\\
\textrm{(k)} &=& \left(\, \frac{5}{2}-\frac{1}{6}\,\pi^2 \,\right)b^3
\, .
\end{eqnarray}
\noindent We come now to graph (a). We employ a method inspired by the
Gegenbauer polynomials technique \cite{Kotikov:Gegenbauer} with the
following differences. First, our space is 2-dimensional, thus the
Gegenbauer expansion reduces to a simple Fourier expansion.
On the other hand, the propagator on the curved space is not simply
$(p^2)^{-\nu}$, but the far more complicated
expression given by (\ref{g(z,z')}). This method can also be useful
to compute higher order diagrams.\\
Leaving apart for a while the combinatorial factor and the factors
coming from the expansion of the exponential of the interaction part,
the integral that gives the tetrahedron graph is
\begin{equation}
\begin{array}{lll}
\vspace{0.4cm}
\hspace{-0.4cm}  \displaystyle{T} &
\hspace{-0.2cm}    = &
\hspace{-0cm} 
\displaystyle{
\int_{\Delta}\frac{d^2 z_2}{\pi b^2\,(1-z_2 \bar{z}_2)^2}\,
\int_{\Delta}\frac{d^2 z_3}{\pi b^2\,(1-z_3 \bar{z}_3)^2}\,
\int_{\Delta}\frac{d^2 z_4}{\pi b^2\,(1-z_4 \bar{z}_4)^2}\;\times}  \\
& &
\vspace{0.3cm}
\hspace{4cm}
\displaystyle{
\times
\;g(z_1,z_2)\,g(z_1,z_3)\,g(z_1,z_4)\,g(z_2,z_3)\,g(z_2,z_4)\,g(z_3,z_4).}  
\end{array}
\end{equation}
Exploiting invariance under $SU(1,1)$ transformations, we can set
$z_1=0$ and $T$ becomes \\
\phantom{------------}
\begin{eqnarray}\label{T}
T =\int_{\Delta}\,\frac{d^2 z_2\hspace{0,2cm}g(|z_2|^2)}{\pi
b^2\,(1-z_2\bar{z}_2)^2}\,  
\int_{\Delta}\,\frac{d^2 z_3\hspace{0,2cm}g(|z_3|^2)}{\pi
b^2\,(1-z_3\bar{z}_3)^2}\,  
\int_{\Delta}\,\frac{d^2 z_4\hspace{0,2cm}g(|z_4|^2)}{\pi
b^2\,(1-z_4\bar{z}_4)^2}\, 
\,g(z_2,z_3)\,g(z_2,z_4)\,g(z_3,z_4)\nonumber \\
\end{eqnarray}
where $g(|z|^2)=g(0,z)$. The Fourier expansion of $g(z,z')$ is given
by 
\begin{equation}
 \label{expansion}
g(z,z')=\sum_{n=0}^{\infty}\,g_n(x,y)\,\cos\beta
\end{equation}
where $\beta$ is the angle between $z$ and $z'$, $x=|z|^2$, $y=|z'|^2$
and the explicit form of $g_n(x,y)$ is
\begin{equation}
 \label{factorization}
g_n(x,y)=  \theta(y-x)\,a_n(x)\,b_n(y)
+\theta(x-y)\,a_n(y)\,b_n(x)
\end{equation} 
with
\[
\begin{array}{ll}
\vspace{0.5cm}
a_0(x) =  \displaystyle{\frac{1+x}{1-x}} 
& \hspace{2cm}
b_0(y) = \displaystyle{ -\,\frac{1}{2}\left(\,\frac{1+y}{1-y}\, \log y \,
+2\,\right)} \\  
a_1(x) = \displaystyle{\frac{\sqrt{x}}{1-x}} 
& \hspace{2cm}
b_1(y)= \displaystyle{\frac{1}{\sqrt{y}}\left(\,\frac{2y}{1-y}\,\log
y+(1+y)\,\right)} \\ 
\end{array}
\]
and, for $n \geqslant 2$,
\[
\begin{array}{ll}
\hspace{-0.17cm}
a_n(x) = \displaystyle{\frac{x^{\frac{n}{2}}}{1-x}}
\left(1-\,\frac{n-1}{n+1}\,x\right)
& \hspace{0.5cm}
b_n(y) =  
\displaystyle{-\,\frac{1}{n\,(n-1)}\,y^{-\frac{n}{2}}
\left(\frac{1+y}{1-y}\,(1-y^n)-n\,(1+y^n)\right)}
\end{array}
\]
obtained by solving the radial equation for the Green function.\\
Performing the angular integrations and exploiting the symmetry of the
Green functions, we can rewrite (\ref{T}) as follows
\begin{eqnarray}
T &=&
 3!\left[\, \int_0^1 dz\,\frac{g(z)}{(1-z)^2}\,
b_0^2(z)\int_0^z dy\,\frac{g(y)}{(1-y)^2}\,
b_0(y)a_0(y)\int_0^y dx\,\frac{g(x)}{(1-x)^2}\,
a_0^2(x) \right.\nonumber \\
& & 
\hspace{0.6cm}
\left.+\,\frac{1}{4}\,\sum_{n=1}^\infty \int_0^1 dz
\frac{g(z)}{(1-z)^2}\,
b_n^2(z)\int_0^z dy \,\frac{g(y)}{(1-y)^2}\,
b_n(y)a_n(y)\int_0^y dx \frac{g(x)}{(1-x)^2}\,
a_n^2(x)\,\right]\nonumber \\
& \equiv & \sum_{n=0}^\infty T_n \hspace{0.1cm}.
\end{eqnarray}
where $z=|z_4|^2$, $y=|z_3|^2$, $x=|z_2|^2$.\\
For $n=0$ and $n=1$, we integrate by parts the last integral, finding
\begin{equation}
\hspace{-2cm}
T_0= \frac{61}{7680}\,-\,\frac{\zeta (3)}{160} 
\qquad 
\hspace{2cm}
T_1=-\,\frac{91}{30720}\,-\,\frac{1}{9600}\,\pi^4\,+\,\frac{7\,\zeta
(3)}{640} \, . 
\end{equation}
For $n \geqslant 2$, we perform an integration by parts in the last
integral with
\begin{equation}
A_n(y) = \int_0^y dx \,\frac{g(x)}{(1-x)^2}\, a_n^2(x) 
\qquad \textrm{and} \qquad
B_n(z) = \int_z^1 dx \,\frac{g(x)}{(1-x)^2}\, b^2_n(x)
\end{equation}
reducing the integrals to the following one dimensional integrals
\begin{equation}
 \label{I_n}
I_n = \int_0^1 dz \, B_n(z) \,\frac{g(z)}{(1-z)^2}\,b_n(z) a_n(z)
A_n(z).
\end{equation}
For $n\geqslant 2$  the primitives of the integrand in
(\ref{I_n}) are very complicated combinations of hypergeometric
functions of high order.  It is more convenient to write the generic
term appearing in the primitives as
\begin{equation}
\int dx \,\frac{x^m}{(1-x)^r}\,\log^p x =\sum_{k=0}^m (-1)^k
\left(\begin{array}{c}
  \hspace{-0.1cm} m \hspace{-0.1cm}\\
  \hspace{-0.1cm} k \hspace{-0.1cm} \end{array}  \right) 
\int dx \,(1-x)^{k-r}\,\log^p x
\end{equation}
and compute separately the finite part of such integrals  since, being
(\ref{I_n}) convergent, the divergent parts cancel out. The algebraic
calculations have been performed with the Mathematica{\tiny$^{TM}$}
 program. We reduce $I_n$ to the form
\begin{eqnarray}
 \label{lastsum}
I_n &=& \frac{1}{240}\,
\left[\,-\,\frac{9+6\,n+2\,n^2}{8\,n\,(n+1)^3}\,-\,
\frac{\pi^2}{6\,n\,(n^2-1)^2}\, \right.
      -\,\frac{n^2}{2\,(n^2-1)}\,\psi_2(n)\,
      -\,\frac{n}{6}\,\psi_3(n)\,   \nonumber \\
& & \hspace{1.2cm}  
     +\,\frac{1}{n\,(n^2-1)^2}\,\psi_1(n)\,
     +\left.\,\frac{1}{24}\,\left(\,n^3\,\psi_3(n)-2-\,\frac{3}{n}\,\right)
\right]
\end{eqnarray}
where $\psi_k(n)$ is the polygamma function, defined as
\cite{polygamma}
\begin{equation}
\psi_k(x)\;=\;(-1)^{k+1}\,k!\,\sum_{r=0}^\infty \,\frac{1}{(r+x)^{k+1}}~.
\end{equation}
At this point, we need to perform the sum of (\ref{lastsum}) for $n$
going from
2 to $\infty$.\\
The first four terms are readly summed. The fifth one can be
decomposed as follows
\begin{equation}
\frac{1}{n\,(n^2-1)^2}\,\psi_1(n)=
\left(\,\frac{1}{n}\,-\,\frac{1/2}{(n-1)}\,-\,\frac{1/2}{(n+1)}\,
      +\,\frac{1/4}{(n-1)^2}\,-\,\frac{1/4}{(n+1)^2}  \,\right)\,\psi_1(n)
\end{equation}
where the first three terms are easily summed, while the last two are
computed exploiting the recursion relations
\begin{equation}\label{recursion}
\psi_k(x+1)\,=\,\psi_k(x)+(-1)^k \,\frac{k!}{x^{k+1}}
\qquad
\hspace{0.7cm}
\psi_k(x-1)\,=\,\psi_k(x)-(-1)^k \,\frac{k!}{(x-1)^{k+1}}~.
\end{equation}
The sum of the last term of (\ref{lastsum}), which is also convergent,
is computed by using the relations
\begin{equation}
2=\lim_{n\rightarrow\infty}\,n^3\,\psi_3(n)
\qquad \hspace{0.5cm}
\textrm{and}
\qquad \hspace{0.5cm}
3=\lim_{n\rightarrow\infty}\,\left[\,n^4\,\psi_3(n)-2\,n\,\right],
\end{equation}
rewriting the limit as a series of differences and using 
(\ref{recursion}).\\
Putting everything together, we find
\begin{equation}
\sum_{n=2}^\infty\,I_n\;=\;-\,\frac{89}{15360}\,+\,\frac{\pi^2}{1536}\,
-\,\frac{\zeta(3)}{1920}~.
\end{equation}
Finally, adding the contribution of $n=0$, $n=1$ and recalling the
factor 6/4 in order to get $T_n$ from $I_n$ when $n\geqslant 2$, we have
\begin{equation}
T\;=\;-\,\frac{19}{5120}\,+\,\frac{\pi^2}{1024}\,-\,\frac{\pi^4}{9600}\,
+\,\frac{\zeta(3)}{256}~.
\end{equation}
Taking into account the combinatorial factor and the factors coming from
the expansion of the exponential of the interaction, we get the 
tetrahedron graph (a). Then, adding the contributions of the other
graphs  
(b), (c), (j) and (k), we find the order $b^3$ in the perturbative
expansion
of $G_3$
\begin{equation}\label{G3}
G_3\;=\;-\,b\,+\,\Big(\,3-2\,\zeta(3)\,\Big)\,b^3\,+\,O(b^5)~.
\end{equation}
To extract $G_3$ from the proposed formula (\ref{U(a)bootstrap}) for 
$U(a)$, we need the perturbative expansion of $\log U(a)$, 
very easy to compute to any order. To order $b^3$, we find
perfect agreement with our perturbative calculation result
(\ref{G3}).\\ 
Further details, along with a comparison with the geometric approach
to Liouville theory \cite{Takhtajan},
will be published elsewhere.

\section*{Acknowledgments}

We are grateful to G. Curci, G. D'Appollonio, E. Remiddi and in
particular to D.  Anselmi and D. Seminara for useful discussions.

\end{document}